\documentstyle[epsfig]{aipproc}
\begin{document}
\title{Electron-positron outflow from black holes and the formation
of winds}
\author{Maurice H.P.M. van Putten}
\address{MIT, Cambridge, MA 02139}
\maketitle

\begin{abstract}
The collapse of young massive stars or the coalescence of
a black hole-neutron star binary is expected to give rise to
a black hole-torus system. When the torus is strongly magnetized,
the black hole produces electron-positron outflow along open  
magnetic field-lines. Through curvature radiation in gaps,
this outflow rapidly develops into a $e^\pm\gamma$-wind, which is
ultra-relativistic and of low comoving density, proposed here as a possible
input to GRB fireball models.
\end{abstract}

Here, I discuss some aspects of
black holes when exposed to external magnetic
fields. For example, black hole-torus systems are a probable outcome 
of the collapse of young massive stars\cite{woo93,pac98}
and the coalescence of black hole-neutron star binaries \cite{pac91},
both of which are possible progenitors of
cosmological gamma-ray bursts (GRBs).
If all black holes are produced by stellar collapse,
they should be nearly maximally rotating
\cite{bar70,bet98}.
A surrounding torus or accretion disk is expected to be magnetized 
by conservation of magnetic flux and linear amplification
(cf. \cite{pac98,klu98}).

A black hole-torus system will have 
open magnetic field-lines from the horizon
to infinity and closed magnetic field-lines between the black hole 
and the torus \cite{mvp99}.
The closed magnetic field-lines 
mediate Maxwell stresses\cite{mvp99b}. This may be seen
by way of similarity to pulsar magnetospheres
\cite{gol69}. In a poloidal cross-section,
the torus can be identified with a pulsar which rotates at an
angular velocity $\Omega_P\sim \Omega_H-\Omega_T$, wherein
the black hole horizon corresponds to infinity.
Then, the inner light-surface \cite{zna77} corresponds
to the pulsar light-cylinder, and
a `bag' attached to the torus to the last closed field-line.
Here, $\Omega_H$ and $\Omega_T$ denote the angular velocities of
the black hole and the torus, respectively.
The work performed by the Maxwell stresses is commonly attributed to
an outgoing Poynting flux emanating from the horizon
\cite{bz77,tho86}.
These Maxwell stresses are likely to be important to the evolution
of the torus, and tend to delay accretion onto the black hole.
The open magnetic field-lines, on the other hand,
enable the black hole to produce an outflow
to infinity. Such outflows generate emissions by deceleration against
the interstellar medium and through internal shocks. 
\mbox{}\\
\mbox{}\\
{\tiny \sffamily To appear in: Explosive Phenom. Astrophys. Compact
Objects, edited by C.-H. Lee, M. Rho, I. Yi \& H.K. Lee, AIP Conf. Proc.}

Here, the outflow along open magnetic field-lines is
studied, and found to produce a pair-dominated 
$e^\pm\gamma$--wind in combination with curvature radiation.

Open field-lines from the horizon to infinity have radiative
ingoing boundary conditions at the horizon
as seen by zero-angular momentum observers (ZAMOs), and 
outgoing boundary conditions 
at infinity.
It is well-known that for an outflow to exist, there must be 
regions in which pairs are created (gaps), somewhere on these
open field-lines \cite{bz77,phi83,pun90,bes00}.
The gaps are powered by an electric current $I$ along the field--lines,
which is limited by a horizon surface resistivity of $4\pi$, in the presence
of a certain potential drop across them. The net particle
flow is limited by the black hole luminosity into the gap. The
magnetosphere within the gaps is differentially rotating, beyond which
the magnetosphere may be force-free and in rigid rotation.
Note that, in contrast, the currents along closed magnetic field-lines
are fixed by the angular velocity $\Omega_T$ of the surrounding matter,
where the gaps are most likely residing between the horizon and the
inner light surface.
Of interest here is the location of the gaps on the open magnetic
field-lines and the power dissipated within, as sites of
linear acceleration of charged particles and their curvature radiation.

A rotating black hole tends to produce electrons and positrons by
spontaneous emission along open magnetic field-lines in an effort
to evolve to a lower energy state by shedding off its angular momentum.
Indeed, in the adiabatic limit, the radiated particles possess a specific 
angular momentum of at least $2M$, whereas the specific angular
momentum of the black hole, $a$, is at most $M$.
In the approximation of an asymptotically uniform magnetic field,
e.g., in a Wald-field \cite{wal74}, the emissions at infinity 
to satisfy a Fermi-Dirac distribution of radiative Landau states, 
neglecting curvature radiation
and magnetic mirror effects. This results from
a modification to the Hawking radiation process\cite{mvp00}.
This is a highly
idealized picture derived in the perturbative limit of small
particle densities, which will be modified significantly by curvature
radiation and the formation of force-free regions.
The spontaneous emission process concerns particles with 
energy-at-infinity $\omega$ below the Fermi-level $V_F$.
Here, $V_F$ is the energy-at-infinity associated with the
particles as seen on a null-generator of the horizon, such as the 
ZAMO-derivative $\xi^a\partial_a=\partial_t-\beta\partial_\phi$,
where $\beta$ denotes the angular velocity of the sky as seen
by ZAMOs. That is,
\begin{eqnarray}
V_F\psi=[\xi^aD_a]^H_\infty\psi=(\nu\Omega_H-eV)\psi,
\end{eqnarray}
where $\Omega_H$ is the angular velocity of the horizon,
using the sign-convention $\psi\propto e^{-i\omega t}e^{i\nu\phi}$.
The energy-at-infinity $\omega$ and
the azimuthal quantum number $\nu$ are associated with
the asymptotically time-like
Killing vector $\partial_t$ and azimuthal Killing vector
$\partial_\phi$, whereas $D_a=i^{-1}\partial_a+eA_a$ denotes the gauge-covariant
derivative in the presence of an electromagnetic vector potential $A_a$.
In calculating $V_F$, it is relevant to identify the 
ground state of the black hole-magnetic field configuration.
It has been shown that the lowest energy state, in the process of
an angular momentum exchange between the black hole and the surrounding
electromagnetic field by variations of the horizon charge $q$,
assumes when $q=2BJ$ \cite{wal74,dok87},
where $J$ is the angular momentum of the black hole. 
Rotation of the equilibrium charge
$q=2BJ$ on the horizon recovers $4\pi BM^2$ as the
maximal horizon flux of the magnetic field from the uncharged flux
$4\pi BM^2\cos\lambda$, where $\sin\lambda=J/M^2$
\cite{dok87,mvp00}.
With the sign
convention that $B$ is parallel to $\Omega_H$, we then have
$V_F=\nu\Omega_H$ with $\nu=eA_\phi$ (for $e^-$) and 
$A_a=B(\partial_\phi)_a/2$ in 
the Wald electrostatic equilibrium state.
Note that the spontaneous emission process is
anti-symmetric under pair-conjugation.

The rate of spontaneous emission is given by a certain 
barrier transmission coefficient in the level-crossing picture
of electrons and positrons \cite{dam77}. This follows 
from frame-dragging by $\beta$, and the resulting shift 
between the energy-at-infinity $\omega$ and the energy 
$\omega_Z$ as seen by ZAMOs:
\begin{eqnarray}
\omega_Z=\pm\sqrt{m_e^2+|eB|(2n+1\pm\alpha)}
=\omega+\nu\beta=\left\{
\begin{array}{cc}
\omega-\nu\Omega_H & \mbox{on the horizon}\\
\omega             & \mbox{at infinity}.
\end{array}\right.
\end{eqnarray}
Here, it is the quantum number $\nu$ which gives rise to
different energies between ZAMOs and Boyer-Linquist observers.
Figure 1 (a) shows an equivalent classical picture, where
the frame-dragging $\beta$ induces a potential energy 
$V_{BL}=e\beta A_\phi$
on a flux surface $A_\phi$=const. with respect to the axis of rotation, 
itself at zero potential in the $q=2BJ$ state. 
\begin{figure}
\epsfig{file=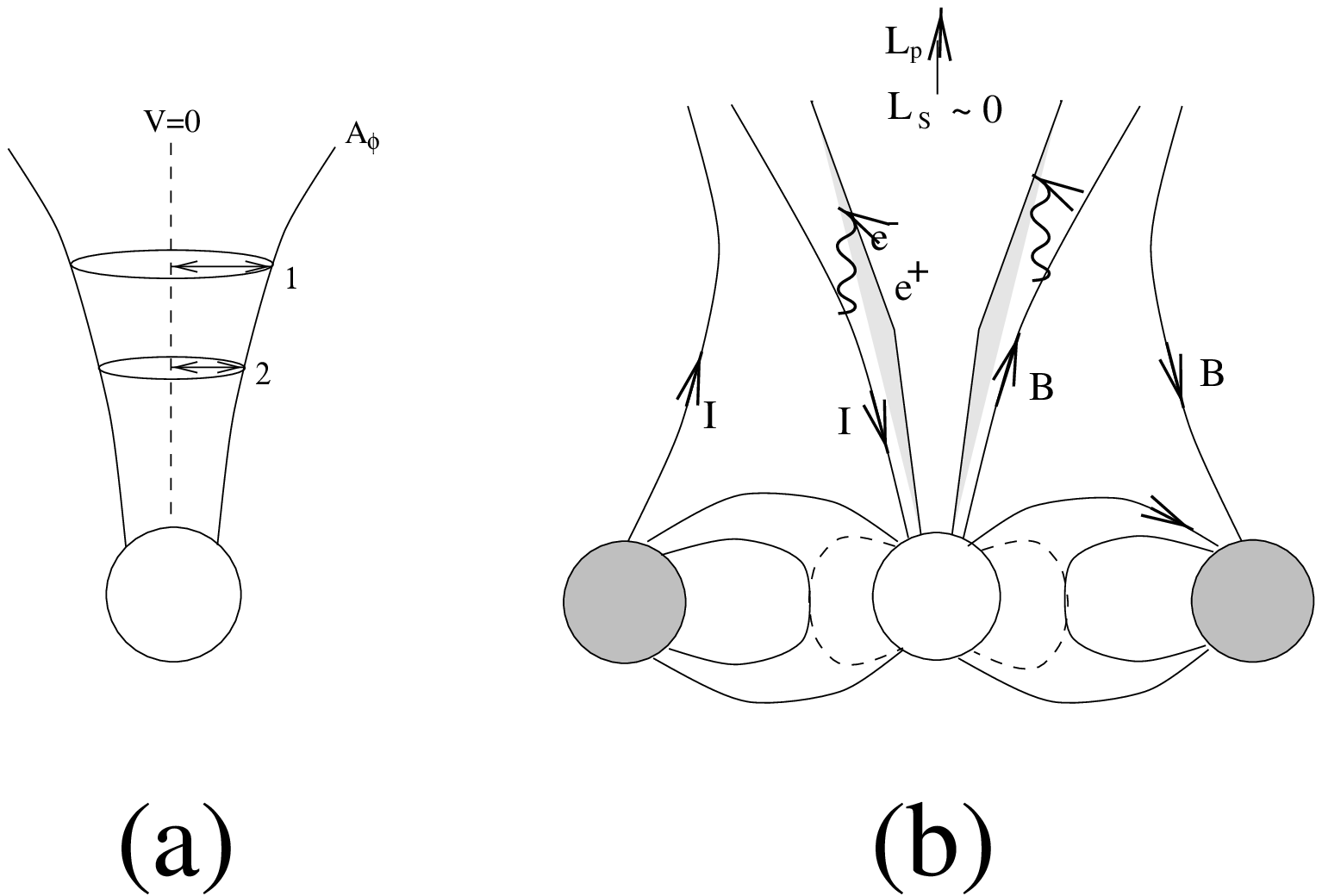,width=140mm,height=100mm}
\mbox{}\\
\mbox{}\\
{\bf Figure 1}. {\small (a) A classical picture of the potential
energy $V_{BL}$ as seen in Boyer-Lindquist coordinates
along surfaces of constant magnetic flux
$A_\phi$. Note that the axis of rotation has zero
potental $V=0$ in electrostatic equilibrium $q=2BJ$.
Hence, $V_{BL}=e\beta A_\phi$ in the presence of frame-dragging
$\beta$. Since $\beta$ describes differential rotation in the
surrounding space-time, a potential drop emerges along $A_\phi=$const.:
$\Delta V_{BL}=\left(\beta_2-\beta_1\right)A_{\phi}$. When
the potential drop is steep, a Schwinger-type process generates
electron--positron pairs.
(b)
Cartoon of 
the formation of a black hole-wind in a black hole-torus system.
There is a minimum opening angle $\theta_{min}\sim\sqrt{B_c/3B}$,
beyond which spontaneous emission along open field-lines
by the black hole is effective.
Flux surfaces with $\theta\sim\theta_{min}$ have a gap length
of order $M$, which decreases for $\theta>\theta_{min}$.
These gaps, indicated in grey,
create pairs, which are subject to linear acceleration
and produce curvature radiation.
The net outflow $L_p$ in particles
is a combination of an inner, current-free outflow 
with vanishingly small Poynting flux $L_S$
inside $\theta<\theta_{min}$
and an outer, current-carrying
outflow with $\theta>\theta_{min}$. Both
derive most of their particles from pair-cascade through
curvature radiation, and flow along open field-lines to 
infinity.
}
\end{figure}
Since $\beta$ describes a differentially rotating space-time, it
varies with distance to the black hole and $V_{BL}$ introduces
a potential energy drop along the magnetic field-lines. When
sufficiently strong, a Schwinger-type process is set in place,
which locally produces pairs at a certain rate per unit volume.
Formally, the rate of pair-production follows from a scattering
calculation in the WKB approximation
\cite{haw75,gib76,dam77} (cf. also
\cite{gol78}).
The pair-production rate is found to be given by a barrier
transmission coefficient $\Gamma\sim e^{-\pi B_c/B\theta^2}$,
where $B_c=m_e^2/e=4.4\times 10^{13}$G is the QED
value of the magnetic field-strength and $\theta$ is the
poloidal angle in Boyer-Linquist coordinates.
More precisely, the gradient
$\eta=-\nabla V_{BL}$ parallel to $B$ drives a pair-production
rate per unit volume by a Schwinger-process
${d^2N}/{dtdV}\sim({e^2\eta B}/{4\pi^2})e^{-\pi B_c/\eta\theta^2}
~~~(B>>B_c,~a\sim M).$ This pair-production process will be in
place, whenever the charge-density is low so that the magnetosphere
remains in differential rotation.

The magnetosphere on open field--lines away from a gap assumes
a force-free, rigidly rotating state with
a Goldreich-Julian charge density\cite{bz77,tho86,tre00}.
This is similar to the analogous case in
pulsar magnetospheres \cite{gol71,gol71b}.
In view of the horizon boundary conditions below,
I shall assume that the gap is attached to the horizon.

To a first approximation, the local structure of a gap follows from
the ingoing radiative horizon--boundary conditions.
The flow in a gap is described by a charge-density $\rho_e$,
a pair-density $n_w$ 
and a Lorentz factor $\Gamma$. This flow is powered by an electric current 
$I$ along the open field--lines to infinity
through a polar cap of area $A_p$ at the cost
of a certain potential drop across.
The ingoing radiative
boundary condition at the horizon applies to electrons
and positrons alike: in the limit as we approach the horizon,
$I$ and the electric charge density $\rho_e$
are no longer independent, but become 
proportional to one another ($cf.$ \cite{pun90}):
\begin{eqnarray}
I\longrightarrow -\rho_e A_p,
\label{EQN_BC}
\end{eqnarray}
since all particles fall into 
the black hole with the velocity of light. The sign
of $\rho_e$ in (\ref{EQN_BC}) is that seen by ZAMOs.
Here, $\rho_e$ (and $n_w$) are normalized by
factoring in the redshift factor. 
$I$ saturates against the horizon surface resistivity of $4\pi$:
$4\pi I\sim \nu\Omega_H$, up to a logarithmic factor on the left hand-side.
Hence, $\rho_e\sim \rho_{GJ}/2$, 
where $\rho_{GJ}=B\Omega_H/2\pi$ is the Goldreich-Julian 
charge density near the horizon.
With curvature $R_B\sim \sqrt{2}M/\theta^2$ of the
Wald--field, curvature radiation produces $n_w>>\rho_e/e$ in
momentum balance: 
$n_{w}2e^2\Gamma^4/3R_B^2\sim \rho_e E_\perp,$
where $E_\perp\sim\nu\Omega_H/Me$ is the equivalent
electric field normal to the horizon as seen
in Boyer-Lindquist coordinates. 
Note, however, that
the magnetic field of the torus will have
larger curvature than that of the Wald-field.
Given energy balance of the outflow
$n_w\Gamma m_e A_{cap}$ with
the total power $IE_\perp L$ dissipated in the gap, where
$L$ is the linear gap size, the solutions are governed by
the unknown $L$.

The gap size $L$ determines the degree to which the black hole
luminosity is put to work in accelerating particles.
The gap produces a radiation pressure $\propto L$,
which acts on the interface with the force-free magnetosphere above.
The interface is probably Raleigh-Taylor
unstable against this radiation pressure.
Moreover, the gap itself may well widen due to this pressure.
The arguments given above are intended as a first sketch
towards the structure of the gaps, and it appears to be of interest 
to consider the gap size $L$ in the context of a detailed stability analysis.

A continuous outflow establishes with
appropriate current closure.
Note that closure over the polar axis introduces
Poynting flux with negative helicity, whereas current
closure over a gap across the equator of the black hole
and the bag of the torus (corresponding to the 
last field-line in pulsar magnetospheres) introduces
Poynting with positive helicity - indicative of positive
energy and angular momentum transport outwards.
As the latter is energetically favorable over the former,
thereby leaving negligible Poynting flux over the axis
of symmetry. A similar conclusion has been found in the case
of current closure around neutron stars
\cite{gol71,gol71b}.
It follows that the black hole-wind is pair-dominated
with the property that $\sigma=L_S/L_p\sim 0$ within
$\theta<\theta_{min}=\sqrt{B_c/3B}$, where
$L_S$ and $L_p$ are the luminosities in Poynting flux 
and pairs, respectively\cite{mvp00b}. Figure 1 (b) sketches this
wind--formation process, assuming a $L$ to be large on the flux
surfaces with $\theta\sim\theta_{min}$.

It will be of interest to look for observational
evidence in GRB--afterglow emissions for 
the presented ultra--relativistic, low density 
pair-dominated wind.
\mbox{}\\
\mbox{}\\
\centerline{\bf Acknowledgements}
\mbox{}\\
This work is partially supported by
NASA Grant 5-7012 and an MIT Reed Award. The author thanks 
the hospitality of Theoretical Astrophysics, Caltech, and the
Korean Institute of Advanced Study (KIAS), where some of this 
work was performed, and gratefully acknowledges stimulating 
discussions with P. Goldreich, E.S. Phinney, R.D. Blandford
and K.S. Thorne.

\end{document}